\documentclass[aps,prb,twocolumn,superscriptaddress]{revtex4-2}
\usepackage{amsmath}
\usepackage{amsfonts,amssymb}
\usepackage{graphicx}
\usepackage{epstopdf}
\usepackage{hyperref}
\usepackage{color}

\begin{document}	
\title{Symmetry classes of dissipative topological insulators with edge dark state}
\author{Fei Yang}
\affiliation{Center for Advanced Quantum Studies, Department of Physics, Beijing Normal University, Beijing 100875, China}
\author{Zheng Wei}
\affiliation{Center for Advanced Quantum Studies, Department of Physics, Beijing Normal University, Beijing 100875, China}
\author{Xianqi Tong}
\affiliation{Center for Advanced Quantum Studies, Department of Physics, Beijing Normal University, Beijing 100875, China}
\author{Kui Cao}
\affiliation{Center for Advanced Quantum Studies, Department of Physics, Beijing Normal University, Beijing 100875, China}
\author{Su-Peng Kou}
\affiliation{Center for Advanced Quantum Studies, Department of Physics, Beijing Normal University, Beijing 100875, China}
\thanks{Corresponding author}
\email{spkou@bnu.edu.cn}
\begin{abstract}
	We classify the dissipative topological insulators (TIs) with edge dark states (EDS) by using the 38-fold way of non-Hermitian systems in this paper.
	The dissipative dynamics of these quadratic open fermionic systems is captured by a non-Hermitian single-particle matrix which contains both the internal dynamics and the dissipation, refereed to as damping matrix $X$.
	And the dark states in these systems are the eigenmodes of $X$ which the eigenvalues' imaginary part vanishes.
	However, there is a constraint on $X$, namely that the modes in which the eigenvalues' imaginary parts are positive are forbidden. 
	In other words, the imaginary line-gap of $X$ is ill-defined, so the topological band theory classifying the dark states can not be applied to $X$.
	To reveal the topological protection of EDS, we propose the double damping matrix $\tilde{X} = \text{diag}\left( X, X^* \right)$, where the imaginary line-gap is well defined. 
	Thus, the 38-fold way can be applied to $\tilde{X}$, and the topological protection of the EDS is uncovered.
	Different from previous studies of EDS in purely dissipative dynamics, the EDS in the dissipative TIs are robust against the inclusion of Hamiltonians. 
	Furthermore, the topological classification of $\tilde{X}$ not only reflects the topological protection of EDS in dissipative TIs but also provides a paradigm to predict the appearance of EDS in other open free fermionic systems.
\end{abstract}
\maketitle
	
\section{Introduction}
Symmetry and topology play the central role in modern physics, which results in many interesting phenomena and future applications, such as robust edge mode in topological insulators (TIs) and topological superconductors (TSs)\cite{topo_Ins1,topo_Ins2,topo_Ins3}.
And the systematic classifications of those phases are explained by the ten-fold way of free fermions, or Altland-Zirnbauer (AZ) symmetry classes\cite{AZ_ori}.
The AZ class provides not only a framework to analyze the topological behavior of system with different symmetries but also gives a paradigm to explore new topological phases.
However, those are for the close Hermitian systems, many physical systems in nature experience dissipation associated with gain and loss, such as atomic, molecular, and optical physics\cite{NH_opt1,NH_opt2,NH_opt3}, the dynamics of these systems are effectively described with a non-hermitian (NH) Hamiltonian.
Dissipation in these NH systems would give rise to many interesting effect that do not have Hermitian counterpart, such as NH skin effects\cite{NH_skin1,NH_skin2,NH_skin3,chiral_damping}, $\mathcal{PT}$-symmetric physics\cite{PT_physics,PT_ep}, and the breakdown of bulk-boundary correspondence \cite{NH_BBC1,NH_BBC2,NH_BBC3,NH_BBC4}.
Moreover, the question that how the dissipation influences the topology of a system attracts much attention, which has been explored in many papers\cite{topo_nonequilibrium1,topo_nonequilibrium2,topo_nonequilibrium3,topo_nonequilibrium4,topo_nonequilibrium5,topo_nonequilibrium6,topo_nonequilibrium7}.
The fundamental interests in studying the topological properties of NH systems are to expand the symmetry classes, which had been settled by Bernard and LeClair based on four fundamental symmetries, resulting in a total of 43 symmetry classes, that is known as Bernard-LeClair class\cite{NH_class}.
While Kawabata {\it et al.} discovered that only 38 of 43 symmetry classes are topologically inequivalent\cite{38-fold}, which is known as 38-fold way.
The 38-fold way provides a paradigm to explore the topology of a NH system, which has been applied in many NH systems, such as the NH Sachdev-Ye-Kitaev Model\cite{NH-SYK}, and symmetry classes of the open quantum system\cite{10_fold_open,topo_open1}.

As we have mentioned, the non-Hermiticity is ubiquitous in nonequilibrium open systems, but most of them are hard to solve especially if the interaction is presented.
Fortunately, the open free fermionic systems are exactly solvable, in which the dissipative dynamics are completely captured by the so called damping matrix $X$, which is a NH single-particle matrix that contains the internal dynamics as well as the dissipation.
The topological phenomena in the dissipative dynamics of quadratic open fermionic systems can be understood with the topological classification of complex spectra of $X$ (or Lindbladian spectra) by using the 38-fold way, in which the symmetries of $X$ play the central role.
The typical future of non-trivial bulk-topology is robust gapless edge-modes. Notably, there are two types of topological edge modes in the Lindbladian spectra (the eigenvalues of $X$, denote as $\{\lambda\}$), the edge zero-frequency modes ($\text{Re}\left(\lambda_{\text{edge}} \right)=0 $) and edge dark states (EDS, $\text{Im}\left(\lambda_{\text{edge}} \right)=0 $).
The edge zero-frequency modes are the dynamical signature of topological order, which forces the damping behavior of the bulk and the edge becomes different\cite{topo_signature}.
While the EDS is usually related to the non-trivial steady-state of system\cite{topo_diss,topo_diss2}.
And the relationship between the topology of $X$ and the edge zero-frequency modes has been uncovered by Lieu {\it et al.} based on the 38-fold way \cite{10_fold_open}, however, the relationship between the topology of $X$ and the EDS is not revealed yet, which is the topic of this paper.

Previous studies of EDS in open free fermionic systems are mainly focused on purely dissipative case, in which the EDS is fragile once the Hamiltonian terms are included\cite{topo_diss,topo_diss2}.
In this paper, we study the EDS in the case of full dynamics, of which both the Hamiltonian and the dissipation are presented. We find that the EDS are protected by the topology of double damping matrix $\tilde{X} = \text{diag}\left( X, X^* \right)$, and it is robust against the including of Hamiltonian.
In our scheme, the artificial degree of freedom $X^*$ has no physical counterpart, it's an auxiliary system which is used to support the imaginary line-gap in $\tilde{X}$. In such way, we can classify $\tilde{X}$ topologically with the 38-fold way and the EDS become the in-gap zero modes of $\tilde{X}$.
Additionally, the symmetry classes of two dissipative TIs of 1D and 2D cases are presented by using the 38-fold way, these examples revealed that the EDS which are protected by the topology of $\tilde{X}$ is robust against the inclusion of Hamiltonian.
It is expected that the double damping matrix that built from  damping matrix of dissipative dynamics in an open free fermionic system can apply to the dissipative TSs as well, which is our future direction.

This paper is organized in the following way. In Sec. II, we introduce the damping dynamics of open free fermionic system and the concepts of dark state in these systems.
In Sec. III, we briefly review the 38-fold way which gives the symmetry classes of NHRM.
In Sec. IV, a double damping matrix which determines the symmetry class of EDS is developed.
In Sec. V, two examples of dissipative TIs with EDS is studied, the dissipative 1D Su-Schrieffer-Heeger (SSH) model and dissipative 2D Qi-Wu-Zhang (QWZ) model. The symmetry classes of those two models are given, and the robustness of EDS is checked. Those results revealed that EDS of dissipative TIs is protected by the topology of the double damping matrix.
We conclude our results and future potential directions in Sec. VI.

\section{damping dynamics of quadratic open fermionic systems}
\subsection{The damping dynamics}
The Liouville dynamics of an open quantum system is usually described by an Lindblad master equation\cite{Master_eq1,Master_eq2,Master_eq3,open_quanutm_sys}
\begin{equation}\label{Master_eq}
	\frac{d}{dt}\rho = -\mathrm{i}[H,\rho]+\sum_\mu\left(2L_\mu^\dagger \rho L_\mu -\{L_\mu^\dagger L_\mu, \rho\}\right),
\end{equation}
that the time evolution of density matrix $\rho$ is governed by two parts, the unitary dynamics and the non-unitary dynamics. The Hamiltonian of system is responsible for the unitary evolution, and the Lindblad operator $L_\mu$ that describes the adding or removing of particles via a Markovian bath is responsible for the non-unitary evolution.

In these open free fermionic system, we consider the Lindblad operators as
\begin{equation}
	L_\mu^g = \sum_sD_{\mu,s}c^\dagger_{\mu,s},\qquad L_\mu^l = \sum_sD_{\mu,s}c_{\mu,s}
\end{equation}
where $\mu$ is the index of the lattice site and $s$ denotes the internal degree of freedom.
And when the pairing term in the Hamiltonian is absent, we can formulate the density matrix with a Gaussian state in terms of one-point correlation function for these quadratic systems\cite{density_corre}
\begin{equation}
	\rho \propto \exp\left(\frac{i}{2}\sum_{m,n}\left[ \ln\frac{C}{\mathbb{I}-C}\right]_{mn}  c^\dagger_m c_n \right),
\end{equation}
where $C$ is single-particle correlation function, $C_{mn} = \text{Tr}(c^\dagger_m c_n \, \rho)$.
Thus the time evolution of the density matrix of a open free fermionic system is fully characterized by it's correlation function
\cite{chiral_damping}
\begin{eqnarray}\label{Time_evolve}
	\mathrm{i}\frac{d}{dt}C &=& [-h^T,C]- \{\mathrm{i}\left( M_g + M_l^T\right) , C\} + 2\mathrm{i}M_g,\nonumber\\
	&=&XC - C X^\dagger + 2\mathrm{i}M_g.
\end{eqnarray}
And $X$ is known as the damping matrix, which is a NH single-particle matrix that contains both the Hamiltonian and the dissipation
\begin{equation}\label{H_effective}
	X=-h^T-\mathrm{i}\left( M_g + M_l^T\right),
\end{equation}
where $H = \sum_{m,n}c^\dagger_ih_{ij}c_j$. 
And the bath matrix $M_g$ and $M_l$ are caused by the dissipation, which are hermitian matrix
\begin{equation}
	\left( M_g\right)_{ij} = \sum_{\mu}D^{g*}_{\mu i}D^g_{\mu j}, \quad \left( M_l\right)_{ij} = \sum_{\mu}D^{l*}_{\mu i}D^l_{\mu j}.
\end{equation}
Damping matrix $X$ provides a complete description of dissipative dynamics, it's becomes more obvious when we consider the speed that an initial state converging to the steady state, i.e. we focus on $\tilde{C}(t)=C(t)-C_{s}$, where $C_{s,ij} = \text{Tr}(c^\dagger_i c_j \, \rho_s)$ is the steady state correlation function.
Then, we find that $\tilde{C}(t)$ is governed by the following equation
\begin{eqnarray}\label{damping_dy}
	\tilde{C}(t) &=& e^{-iXt} \cdot \tilde{C}_0 \cdot e^{-iX^\dagger t},\nonumber\\
	&=&\sum_{m,n}\mathrm{e}^{\mathrm{i}\left( -\lambda_n + \lambda^*_m \right)t }|u^R_n\rangle\langle u^L_n|\tilde{C}_0 |u^L_m\rangle\langle u^R_m|.
\end{eqnarray}
The second step in Eq.(\ref{damping_dy}) is obtained using eigen-decomposition method, where $\{\lambda_n\}$ are the eigenvalues of $X$, that is Lindbladian spectra of the system,
and $|u^R_n\rangle$ and $|u^L_n\rangle$ satisfy biorthogonal condition, $\langle u^L_n|u^R_n\rangle=\delta_{mn}$, $X|u^R_n\rangle = \lambda_n|u^R_n\rangle$, $X^\dagger|u^L_n\rangle = \lambda^*_n|u^L_n\rangle$.
It's obvious that Eq.(\ref{damping_dy}) is coincide with the Schr\"{o}dinger equation in quantum mechanics, of which the dynamic generator is a NH single-particle matrix, and $\tilde{C}(t)$ is analogous to the density matrix.
So the topological property of the dissipative dynamics is captured by the NH damping matrix $X$.
The same dissipative dynamics described by the damping matrix when the pairing-term is included in the open free fermionic system by using the method of third quantization\cite{Third_quantization1,Third_quantization2}, while the quansi-particle is changed into Marjorana fermion.

\subsection{Lindbladian spectra and the dark state}
\begin{figure}[h]
	\includegraphics[width=0.3\textwidth]{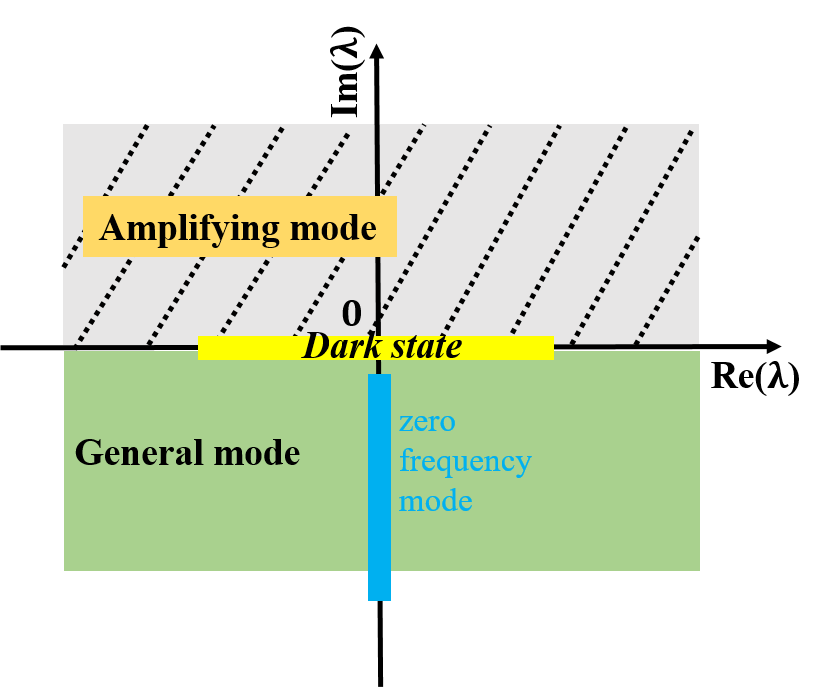}
	\caption{The lindbladian spectra of open free fermionic system in the complex plane. Where the physical modes are those $\text{Im}\left(\lambda \right)<0 $ (green), the dark states are those $\text{Im}\left(\lambda \right)=0 $ (yellow), and the edge zero frequency modes are those $\text{Re}\left(\lambda \right)=0 $ (light blue). However, The modes that $\text{Im}\left(\lambda \right)>0 $ (gray) are the amplifying mode, which is unphysical.}
	\label{Lindblad_spectra}
\end{figure}

In the dissipative dynamics, the dynamic information is hidden in the Lindbladian spectra (or rapidity spectra), which is the eigenvalues' spectra of damping matrix $X$ in the complex plane (denote as $\{\lambda\}$) in the quadratic open free fermionic system, as seen in Fig.\ref{Lindblad_spectra}.
In which the imaginary parts of $\lambda$ specify the speed that the initial state converges to the steady state, the smaller the $\text{Im}\left(\lambda \right)$ the quicker it converges to the steady state.
Such that the spectral gap is defined as $\Delta = 2\max\left[ \text{Im}\left(\lambda \right)\right] $\cite{Third_quantization1,topo_diss2}.
In the Lindblidian spectra, the general modes are those with negative imaginary parts ($\text{Im}\left(\lambda \right)<0 $),  which decays over time.
And if there only have general modes in the spectra, then there is a unique steady state of the Liouville dynamics\cite{Third_quantization1,unique_so}.
The modes with positive imaginary part are forbidden ($\text{Im}\left(\lambda \right)>0 $), which are unphysical because they are amplified over time.
The dark state is the eigenmode which it's eigenvalue's imaginary part vanishes ($\text{Im}\left(\lambda \right)=0 $), these modes are neither decay nor amplify over time. 
Because the dark state is decoupled from the dissipative dynamics, so it is also a steady-state of the system governed by Eq.(\ref{Master_eq}), such that the dark state implies the non-unique steady state of Liouville dynamics\cite{Third_quantization1}.
And the zero frequency mode is the eigenmode which it's eigenvalue's real part vanishes ($\text{Re}\left( \lambda \right)=0 $).

Similar to the gapless edge mode which protected by the topology of Hamiltonian, EDS and edge zero-frequency mode in the dissipative dynamics are related to the topology of damping matrix.
From the point view topological band theory, these edge modes are in-gap states.
There are three kinds of bulk-gap in the complex spectra, the point-gap, real line-gap, and imaginary line-gap, which support different kinds of in-gap edge modes.
The closing of point gap are those $\lambda=0$, and the closing of real (imaginary) line-gap are those $\text{Re}\left(\lambda \right)=0 $ ($\text{Im}\left(\lambda \right)>0 $).
So the topological classification of dissipative dynamics is to study the complex spectra of damping matrix.
S. Lieu and {\it et al.} firstly apply such principle in the open free fermionic system, and classify the dissipative dynamics with edge zero-frequency mode by using the 38-fold way, which leads to the Ten-fold way for quadratic lindbladians, denote as LMC class\cite{10_fold_open}.
However, it's not known whether the dissipative dynamics with EDS can be classified with a similar scheme.

\section{topological classification of Non-hermitian random matrix}
There are 43 non-equivalent symmetry classes of NHRM, which is known as Bernard-LeClair class \cite{NH_class}.
The topological classification of those NHRM based on the AZ scheme is proposed by K. Kawabata and {\it et al.}, that there only have 38 of topological inequivalent symmetry classes, which is the 38-fold way \cite{38-fold}.
The main principle of the classification is to flatten the spectra of a NH matrix, it is accomplished by the unitary flatten of point-gap, hermitian flatten of real line-gap and anti-hermitian flatten of imaginary line-gap.
These flatten procedures keep the symmetries and the bulk-gap of complex spectra, such that the topological classification is identical to it's hermitian (or anti-hermitian) counterpart, in which the principle in the AZ scheme is used.
There are three fundamental symmetries in the AZ class, time-reversal symmetry (TRS), particle-hole symmetry (PHS), chiral symmetry (CS),
\begin{eqnarray}	
	\text{TRS}:&& \qquad U_t\cdot H^*\cdot U^{-1}_t = H,\\	
	\text{PHS}:&& \qquad U_c \cdot H^T \cdot U^{-1}_c = - H,\\
	\text{CS}:&& \qquad S \cdot H^\dagger \cdot S^{-1} = - H,
\end{eqnarray}
where $U_t$ and $U_c$ are unitary matrices, and square to 1 or to -1, i.e., we have $U_{c,t}U^*_{c,t}=\pm1$.
And CS is a combination of TRS and PHS, such that $S = U_t U^*_c$.
So we have $3 \times 3 = 9$ kinds of symmetry classes, and another symmetry class is that only the CS is satisfied, which gives total of 10 symmetry classes.
In contrast to the Hermitian case, there are a variant of TRS and PHS for NH matrix, which is defined as
\begin{eqnarray}	
	\text{TRS}^\dagger:&& \qquad U_t\cdot H^T\cdot U^{-1}_t = H,\\	
	\text{PHS}^\dagger:&& \qquad U_c \cdot H^* \cdot U^{-1}_c = - H.
\end{eqnarray}
These symmetry class is denoted as AZ$^\dagger$ class.
Compare to AZ class, there only have 6 independent symmetry classes.
Furthermore, there is an additional symmetry for NH matrix, the sub-lattice symmetry (SLS), 
\begin{equation}
	U_s \cdot  H \cdot U^{-1}_s = -H, \qquad U_s^2 = 1,
\end{equation}
which is equivalent to CS for Hermitian random matrix, while it is an additional symmetry of NH matrix since $H^\dagger\neq H$.
Such additional second-order symmetry would alter the classification space\cite{second_order}, then symmetry classes is enriched for NHRM, and gives another 22 symmetry classes that specified by the commutation/anti-commutation relations of $U_s$ with TRS or PHS.
And results, we have $10+6+22=38$ symmetry classes, this is the 38-fold way of NHRM.
Another internal symmetry of NHRM is the pseudo-hermitian \cite{p_hermiticity1,p_hermiticity2}, $\eta H^\dagger \eta^{-1}=H$, which is a second-order symmetry that gives the same symmetry classes as SLS.

\section{The Double damping matrix and the system class of EDS}
In the topological band theory, if there are gapless modes that closing gap at the edge of system, then the bulk-gap which has non-trivial topology is unavoidable.
Which means that if the EDS in the dissipative dynamics are protected by the topology of system, then we must have two bands in the system of which the imaginary parts with opposite sign, one is negative, the other one is positive.
\begin{figure}[h]
	\includegraphics[width=0.3\textwidth]{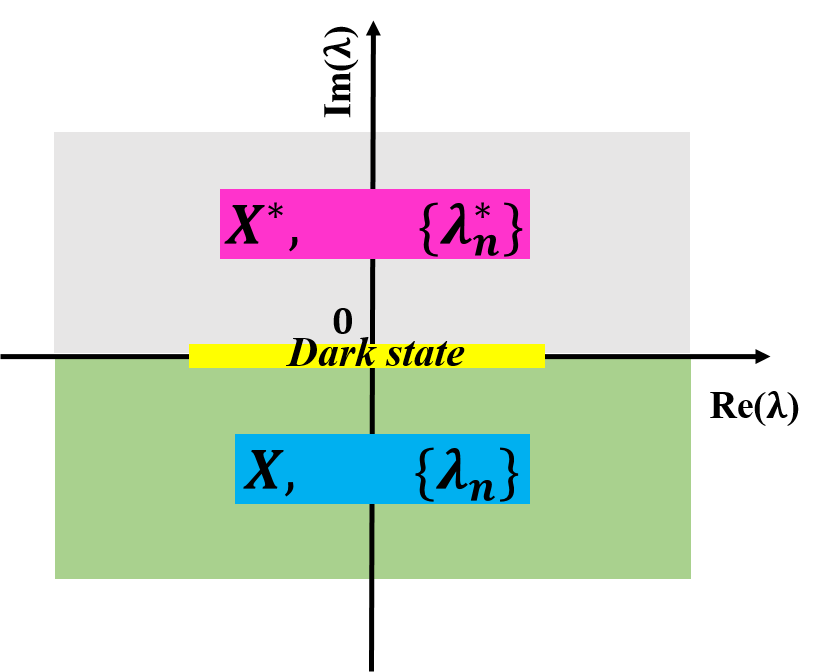}
	\caption{The double Lindbladian spectra of open free fermionic system, that the modes which $\text{Im}(\lambda_n)>0$ and $\text{Im}(\lambda_n)<0$ are allowed. However, the modes which $\text{Im}(\lambda_n)>0$ are unphysical, the including of those modes is only for the convenience of symmetry classification.}
	\label{Double_spectra}
\end{figure}

However, the eigenmodes that $\text{Im}(\lambda_n)>0$ is forbidden in the damping matrix, so we can't define the imaginary line-gap.
Which means that the dark states of an open quantum system might have no topological protection.
To reveal the topological protection of EDS, we combine $X$ with it's complex conjugate $X^*$ to form a double damping matrix
\begin{equation}
	\tilde{X}=\begin{pmatrix}
		X & 0\\
		0 & X^*
	\end{pmatrix}.
\end{equation}
That the redundant freedom $X^*$ is only for the convenience of topological classification, it is unphysical and should be discarded in the dissipative dynamics.
Both the positive and the negative imaginary parts are present in the spectra of $\tilde{X}$, as seen in Fig.\ref{Double_spectra}, so the imaginary line-gap is well defined in $\tilde{X}$.
Such that nontrivial topology of $\tilde{X}$ would indicates the appearance of in-gap edge-mode that closing the imaginary line-gap, which is the EDS of dissipative dynamics.

Therefore, if $\text{Im}\left(\lambda \right)\neq0$, then the imaginary line-gap of non-hermitian matrix $\tilde{X}$ can be defined as $2\max\left[ \text{Im}\left(\lambda \right)\right]$, which is the spectra gap of Liouville dynamics.
And the corresponding symmetry class is identified with the symmetries that $\tilde{X}$ obeys.
We find that TRS is automatically satisfied in $\tilde{X}$, $U_t\tilde{X}^*U_t^{-1} = \tilde{X}$, where $U_t=\tau_x\otimes \mathbb{I} $ or $\tau_y\otimes \mathbb{I} $, and $\tau_x,\tau_y$ is acting on the artificial degree of freedom.
However, both $\tau_x\otimes \mathbb{I}$ and $\tau_y\otimes \mathbb{I}$ are allowed for $\tilde{X}$, which means that the symmetry classes of $\tilde{X}$ is non-unique.
Unfortunately, the ambiguity in the symmetry classes can't be removed, the interesting thing is that the robustness of EDS against the perturbations is enhanced if the symmetry classes which give non-trivial topology is non-unique.
The robustness of EDS can be checked by considering the perturbation which corresponds to the coupling between the degrees of freedom that belongs to $X^*$ and $X$.
If the coupling that satisfy the symmetry constraint can gap out the EDS, then the corresponding topology is trivial, otherwise, the topology is non-trivial.
To determine the robustness of EDS, we study the coupled damping matrix
\begin{equation}
	\tilde{X}_c=\begin{pmatrix}
		X & C\\
		C^\dagger & X^*
	\end{pmatrix}.
\end{equation}
The coupling $C$ can breaks or preserves the symmetry of $\tilde{X}$, this is depends on the symmetry operation. For example, if $C=C^T$, the TRS is preserved if $U_t = \tau_x\otimes\mathbb{I}$, while the TRS is broken if $U_t = \tau_y\otimes\mathbb{I}$. On the contrary, $C=-C^T$ preserve TRS for $U_t = \tau_y\otimes\mathbb{I}$ and break TRS for $U_t = \tau_x\otimes\mathbb{I}$.
Consequently, if the symmetry classes of $\tilde{X}$ corresponds to the symmetry operation $\tau_x\otimes \mathbb{I}$ or $\tau_y\otimes \mathbb{I}$ is topological non-trivial in a given spatial dimension, then one can check that whether the EDS are exist or not in the spectra of $\tilde{X}_c$ to verify the robustness of EDS.
If the EDS still present in the spectra of $\tilde{X}_c$, that is to say the the EDS can't be removed by the perturbation which respect the symmetry constraint, therefore topological protection of EDS is proved.
Instead, if the EDS is disappear in the spectra of $\tilde{X}_c$, then the corresponding topology is trivial, and the EDS is a trivial mode of dissipative dynamics.
In this paper, we considering the perturbation which preserve the symmetry as $C_1 = c\cdot\tau_x \otimes \sigma_x$ for $U_t = \tau_x\otimes\mathbb{I}$ and $C_2 = c\cdot\tau_x \otimes \sigma_y$ for $U_t = \tau_y\otimes\mathbb{I}$.

In a word, the topological classification of double damping matrix $\tilde{X} = \text{diag}\left( X, X^* \right)$ can reveal the topology of the dissipative dynamics with EDS.
If the topology of $\tilde{X}$ is non-trivial in a given spatial dimension, then there is EDS in the corresponding open quantum system which described by $X$.
And the ambiguities in the symmetry classes of $\tilde{X}$ can enhance the robustness of EDS against the perturbation.

\section{Some Examples}
\subsection{dissipative SSH model}
The SSH model describes spinless fermions hopping on a one-dimensional (1D) lattice with staggered hopping amplitudes, and it's Bloch Hamiltonian is written as\cite{topo_insulators_lecture1,topo_insulators_lecture2}
\begin{equation}
	H_{\text{ssh}}(k) = (v+w\cos k)\, \sigma_x + w\sin k\,\sigma_y.
\end{equation}
Where $\sigma_{x,y,z}$ represents the internal degree of freedom, and denote as sub-lattice $A$ and $B$.
In order to have EDS, we consider the dissipators per lattice site as
\begin{equation}
L^g_\mu = \sqrt{\gamma_g}c^\dagger_{\mu,A},\qquad L^l_\mu = \sqrt{\gamma_l}c^\dagger_{\mu,A},
\end{equation}
where subscript $\mu$ represents the index of lattice site.
With simple derivation, we get damping matrix of the dissipative SSH model as
\begin{equation}
	X_{\text{ssh}}(k) = -\left( v+w\cos k\right)\sigma_x + w\sin k \, \sigma_y- \mathrm{i}\gamma\, \sigma_z-i\gamma\mathbb{I},
\end{equation}
where $\gamma = \frac{\gamma_g+\gamma_l}{2}$. The spectrum of $X_{\text{ssh}}(k)$ is
\begin{equation}
	\lambda_{\text{ssh}}(k) = -i\gamma\pm\sqrt{v^2+w^2+2vw\,\cos\,k - \gamma^2}.
\end{equation}
It's obvious that $\text{Im}\left( \lambda_{\text{ssh}}\right) \leq 0$, and "=" is only for the real gap closing point where $v=w$.
In other words, the closing of imaginary line-gap is accompanied by the closing of real gap for $\tilde{X}_{\text{ssh}}(k)$, which means the edge modes of $\tilde{X}_{\text{ssh}}(k)$ are those that closing the point-gap.
Furthermore, if the winding number of $H_{\text{ssh}}$ is non-trivial in the bulk, then the point-gap is closed at the edge of system. So that the critical point of $X_{\text{ssh}}(k)$ to has EDS is identical to the topological phase transition point of $H_{\text{ssh}}$, which means that the EDS are the dynamic signature of the topological order of internal dynamics.

Next, we see the topological classification of dissipative SSH model with EDS. The symmetries of $X_{\text{ssh}}(k)$ are as follows
\begin{eqnarray}
	\text{PHS}^\dagger:&&\qquad \sigma_z X_{\text{ssh}}^*(k) \sigma_z  = -X_{\text{ssh}}(-k),\\
	\text{TRS}^\dagger:&&\qquad X_{\text{ssh}}^T(k) = X_{\text{ssh}}(-k),
\end{eqnarray}
then, $X_{\text{ssh}}(k)$ is belongs to the class BDI$^\dagger$ with a $\mathbb{Z}$ classification from LMC classes in the 1D case, which means there are edge-modes that closing the real line-gap ($\text{Re}\left(\lambda_{\text{edge}} \right)=0 $).
However, one of such edge-mode is the EDS ($\text{Im}\left(\lambda_{\text{edge}} \right)=0 $), that can't be specified by the class BDI$^\dagger$.
To reveal the topological protection of EDS in the dissipative SSH model, we considering the double damping matrix, which has the following symmetry
\begin{eqnarray}
	\text{SLS}: &&\qquad U_s \cdot \tilde{X}_{\text{ssh}}(k)\cdot U^{-1}_s  = -\tilde{X}_{\text{ssh}}(k),\\
	\text{PHS}:&& \qquad U_c \cdot \tilde{X}_{\text{ssh}}^T(k) \cdot U^{-1}_c =- \tilde{X}_{\text{ssh}}(-k),\\
	\text{TRS}:&& \qquad U_t\cdot \tilde{X}_{\text{ssh}}^*(k)\cdot U^{-1}_t = \tilde{X}_{\text{ssh}}(-k),
\end{eqnarray}
where $U_c = U_s = \tau_x \otimes \sigma_z$ or $\tau_y \otimes \sigma_z$, and $U_t = \tau_x \otimes \mathbb{I}$ or $\tau_y \otimes \mathbb{I}$.
The ambiguities of symmetry operation leads to four possible symmetry classes, which is obtained with the permutation of $U_s$ and $U_t$, as seen in Table.I.
\begin{table}[h]
	\label{ssh_class}
	\begin{center}
		\begin{tabular}{ c| c c }
			\hline
			\hline
			{$U_t$} $\setminus$ {$U_s$} & $\tau_x\otimes\sigma_z$  & $\tau_y\otimes\sigma_z$ \\
			\hline
			$\tau_x\otimes\mathbb{I}$& $\mathcal{S}_{++}$\,BDI &  $\mathcal{S}_{-+}$\,CI \\			
			$ \tau_y\otimes\mathbb{I}$&  $\mathcal{S}_{-+}$\,DIII &  $\mathcal{S}_{++}$\,CII \\			
			\hline
			\hline
		\end{tabular}
		\caption{The symmetry classes of $\tilde{X}_{\text{ssh}}(k)$ with sub-lattice symmetry (SLS), where the first and second subscripts of $\mathcal{S}_{\pm\pm}$ specifies the commutation ($+$) and anti-commutation ($-$) relation to the time-reversal symmetry (TRS) and particle-hole symmetry (PHS) correspondingly. That $U_sU_t = \pm \epsilon_t U_t U^*_s$, $U_sU_c = \pm \epsilon_c U_c U^*_s$, where $U_tU^*_t = \epsilon_t$ and $U_cU^*_c = \epsilon_s$.}
	\end{center}
\end{table}

\begin{figure}[h]
	\includegraphics[width=0.5\textwidth]{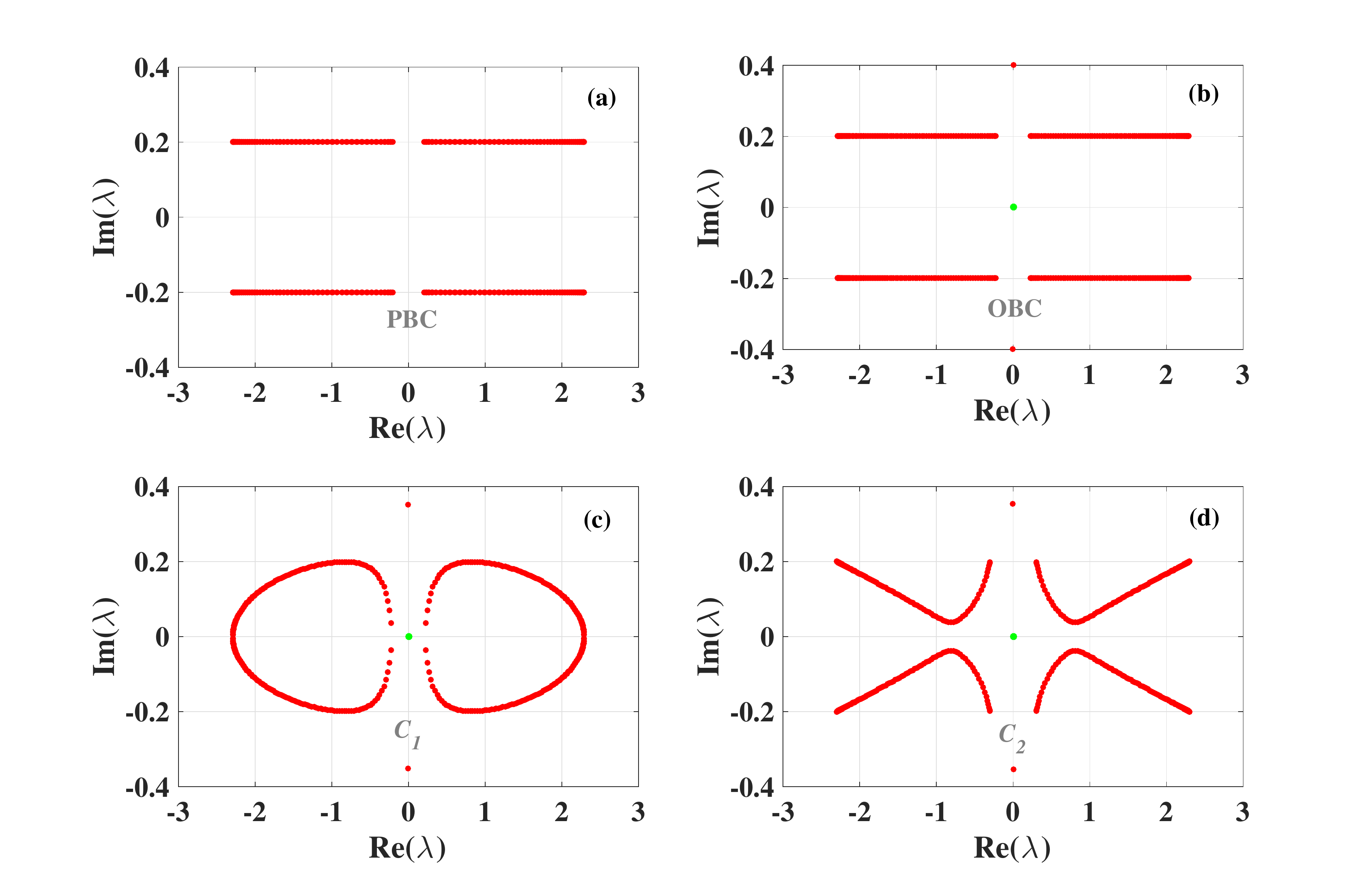}
	\caption{The double Lindbladian spectra of dissipative SSH model with edge dark state (EDS) in periodic boundary condition (a, PBC) and open boundary condition (b, OBC) for $L=100$, the EDS is marked in green. Where $v=1$, $w=1.3$, and $\gamma_g=\gamma_l=0.2$. And the spectra of coupled damping matrix $\tilde{X}_{c,\text{ssh}}$ for $C_1$ (c) and $C_2$ (d), where $c = 0.2$, it is obvious that EDS is robust both to the $C_1$ and $C_2$.}
	\label{SSH_Cspectra}
\end{figure}

When the symmetry classes are identified, then the corresponding classifying space with the point-gap and it's topological properties at a certain spatial dimension is revealed, as seen in Table.II. 
In the 1D case, the model falls into class $\mathcal{S}_{++}$BDI with $\mathbb{Z}$ classification or $\mathcal{S}_{++}$CII with $2\mathbb{Z}$ classification.
The label $2\mathbb{Z}$ indicates the topological number is an even integer, which is isomorphic to the $\mathbb{Z}$ classification, in a word, the topological number of dissipative SSH model with EDS is characterize by an integer $\mathbb{Z}$.
This can be understood with the symmetry analysis, the coulpling $C_1$ respects TRS and PHS of class $\mathcal{S}_{++}$BDI, and the coulpling $C_2$ respects TRS and PHS of class $\mathcal{S}_{++}$CII, which means that the EDS in $\tilde{X}_{\text{ssh}}$ are robust both to the coupling $C_1$ and $C_2$, which can be seen in Fig.\ref{SSH_Cspectra}(c,d).
However, the couplings $C_1$ and $C_2$ break the symmetries of class $\mathcal{S}_{-+}$DIII and class $\mathcal{S}_{-+}$CI, then gives trivial topological classification.

\begin{table}[h]
	\label{ssh_classify}
	\begin{center}
		\begin{tabular}{  c | c c c c c}
			\hline
			\hline
			& classifying space & $d=0$ & $d=1$ & $d=2$ & $d=3$\\
			\hline
			$\mathcal{S}_{++}$\,BDI &  $\mathcal{R}_1$ & $\mathbb{Z}_2$ & $\mathbb{Z}$& 0 & 0 \\			
			$\mathcal{S}_{-+}$\,CI &  $\mathcal{C}_0$  &$\mathbb{Z}$ & 0 & $\mathbb{Z}$  &0\\			
			$\mathcal{S}_{-+}$\,DIII & $\mathcal{C}_0$   & $\mathbb{Z}$ & 0& $\mathbb{Z}$ &0\\
			$\mathcal{S}_{++}$\,CII &  $\mathcal{R}_5$ &0 & $2\mathbb{Z}$& 0 & $\mathbb{Z}_2$\\
			\hline
			\hline
		\end{tabular}
		\caption{The classifying spaces and the topological numbers of symmetry classes of $\tilde{X}_{\text{ssh}}(k)$ with point-gap, where $d$ is the number of dimensions.}
	\end{center}
\end{table}

The Lindbladian spectra of dissipative SSH mode in the periodic boundary condition (PBC) and open boundary condition (OBC) are presented in Fig.\ref{SSH_Cspectra}, as well as coupled damping matrix $\tilde{X}_{c,\text{ssh}}$.
The EDS is robust both to the coupling $C_1$ and $C_2$, this is because that both class $\mathcal{S}_{++}$CII and class $\mathcal{S}_{++}$BDI can characterize the topological protection of EDS, as seen in Fig.\ref{SSH_Cspectra}(c,d).

\subsection{dissipative QWZ model}
The QWZ model is a two-dimensional (2D) Chern insulator that describes the spinless fermions hopping in 2D lattice.
The degree of freedoms (the orbital) for each unit cell are 2, which we denote them as sub-lattice $A$ and $B$, then Bloch Hamiltonian of QWZ model is \cite{topo_insulators_lecture1,topo_insulators_lecture2}
\begin{equation}\label{qwz_bloch}
	H_{\text{qwz}}(\mathbf{k}) = \sin k_x \,\sigma_x + \sin k_y \,\sigma_y + \left( u+\cos k_x + \cos k_y\right)\sigma_z,
\end{equation}
Similarly, $\sigma_{x,y,z}$ represents the internal degree of freedom.
We consider the following dissipators per lattice site
\begin{equation}
	L^g_\mu = \sqrt{\gamma_g}\left( c^\dagger_{\mu,A} + ic^\dagger_{\mu,B} \right) ,\qquad L^l_\mu = \sqrt{\gamma_l}\left( c_{\mu,A} - ic_{\mu,B} \right),
\end{equation}
where subscript $\mu$ respresents the index of lattice.
Then the damping matrix of dissipative QWZ model is
\begin{eqnarray}\label{QWZ_damp}
	X_{\text{qwz}}(\mathbf{k}) = \sin k_x \,\sigma_x + \left( \sin k_y -i\gamma \right) \, \sigma_y\nonumber\\
	 - \left( u+\cos k_x  + \cos k_y \right) \sigma_z - i\gamma\mathbb{I},
\end{eqnarray}
and the spectrum of $X_{\text{qwz}}(\mathbf{k})$ is obtained as
\begin{eqnarray}
	&&E_{\text{qwz}}(\mathbf{k}) = -i\gamma\\
	&&\pm\sqrt{\left(  u+\cos k_x  + \cos k_y \right)^2+\sin^2k_x + \left(\sin k_y -i\gamma \right)^2 },\nonumber
\end{eqnarray}
it's obvious that $\text{Im}\left( E_{\text{qwz}}\right) \leq 0$, and "=" is only that $\left(  u+\cos k_x  + \cos k_y \right)^2+\sin^2k_x=0$, which is also the gap closing points where $u=-2,0,2$.
Amazingly, the closing of imaginary line-gap can also be satisfied at the edge of system when the Chern number of $H_{\text{qwz}}$ is non-zero.
So, the EDS of dissipative QWZ model ware the dynamic signature of the topological order of internal dynamics.

\begin{table}[h]
	\label{qwz_class}
	\begin{center}
		\begin{tabular}{ c| c c }
			\hline
			\hline
			{$U_t$} $\setminus$ {$U_s$} & $\tau_x\otimes\sigma_x$  & $\tau_y\otimes\sigma_x$ \\
			\hline
			$\tau_x\otimes\mathbb{I}$& $\mathcal{S}_{+}$\,AI &  $\mathcal{S}_{-}$\,AI \\			
			$ \tau_y\otimes\mathbb{I}$&  $\mathcal{S}_{-}$\,AII &  $\mathcal{S}_{+}$\,AII \\			
			\hline
			\hline
		\end{tabular}
		\caption{The symmetry classes of $\tilde{X}_{\text{qwz}}(\mathbf{k})$ with sub-lattice symmetry (SLS) $\mathcal{S}$, where subscripts of $\mathcal{S}_{\pm}$ specifies the commutation ($+$) and anti-commutation ($-$) relation to the time-reversal symmetry (TRS). That $U_sU_t = \pm \epsilon_t U_t U^*_s$, where $U_tU^*_t = \epsilon_t$.}
	\end{center}
\end{table}

Then, we study the topological protection of EDS. For $X_{\text{qwz}}$, it's symmetry is 
\begin{equation}
	\text{PHS}^\dagger: \qquad\sigma_x X_{\text{qwz}}^*(-\mathbf{k}) \sigma_x  = -X_{\text{qwz}}(\mathbf{k}),
\end{equation}
In the 2D case, it belongs to class D$^\dagger$ with $\mathbb{Z}$ classification in the LMC class, which means there are edge-modes that closing the real line-gap.
However, the EDS in the dissipative QWZ model can't be specified by class D$^\dagger$.
The symmetry protection of EDS can be revealed in the double damping matrix, $\tilde{X}_{\text{qwz}}(\mathbf{k})$ has the following symmetry
\begin{eqnarray}
	\text{SLS}: &&\qquad  U_s\cdot \tilde{X}_{\text{qwz}}(\mathbf{k})\cdot U^{-1}_s  = -\tilde{X}_{\text{qwz}}(\mathbf{k}),\\
	\text{TRS}:&& \qquad U_t\cdot\tilde{X}_{\text{qwz}}^*(\mathbf{k})\cdot U^{-1}_t = \tilde{X}_{\text{qwz}}(-\mathbf{k}), 
\end{eqnarray}
where $U_s = \tau_x \otimes \sigma_x$ or $\tau_y \otimes \sigma_x$, and $U_t = \tau_x \otimes \mathbb{I}$ or $\tau_y \otimes \mathbb{I}$.
Identical to the SSH model, the ambiguities of symmetry operation leads to four possible symmetry classes, as seen in Table.III

\begin{table}[h]
	\label{qwz_classify}
	\begin{center}
		\begin{tabular}{  c | c c c c c}
			\hline
			\hline
			& classifying space & $d=0$ & $d=1$ & $d=2$ & $d=3$\\
			\hline
			$\mathcal{S}_{+}$\,AI &  $\mathcal{R}_1$ & $\mathbb{Z}_2$ & $\mathbb{Z}$& 0 & 0 \\			
			$\mathcal{S}_{-}$\,AI &  $\mathcal{R}_3$  &0 & $\mathbb{Z}_2$ & $\mathbb{Z}_2$  &$\mathbb{Z}$\\			
			$\mathcal{S}_{-}$\,AII & $\mathcal{R}_7$   & 0 & 0& 0 &$2\mathbb{Z}$\\
			$\mathcal{S}_{+}$\,AII &  $\mathcal{R}_5$ &0 & $2\mathbb{Z}$& 0 & $\mathbb{Z}_2$\\
			\hline
			\hline
		\end{tabular}
		\caption{The classifying spaces and the topological numbers of symmetry classes of $\tilde{X}_{\text{qwz}}(\mathbf{k})$ with imaginary line-gap, where $d$ is the number of dimensions.}
	\end{center}
\end{table}

\begin{figure}[h]
	\includegraphics[width=0.5\textwidth]{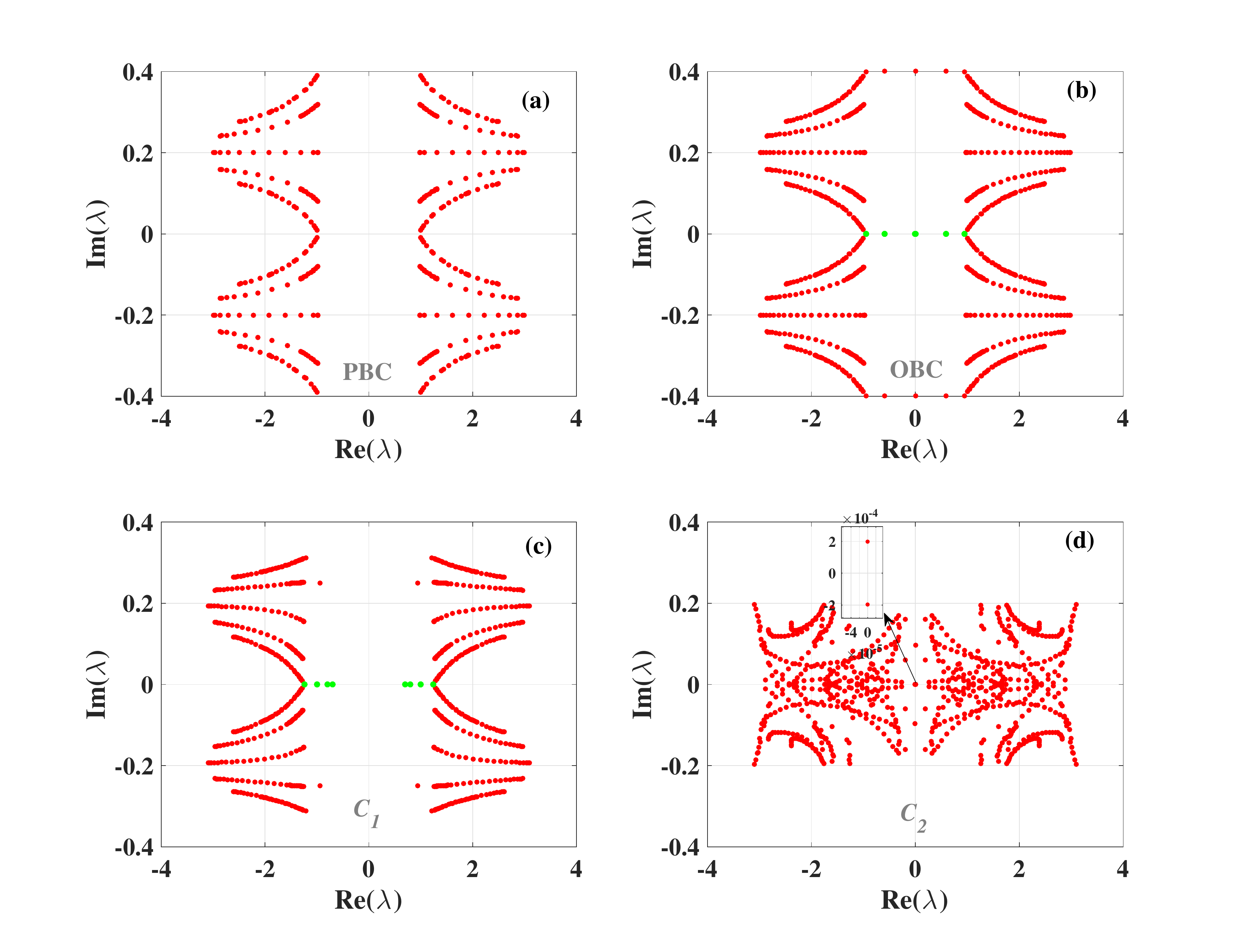}
	\caption{The double Lindbladian spectra of dissipative QWZ model with edge dark state (EDS) in periodic boundary condition (a, PBC) and open boundary condition (b, OBC) for $L_x = 20$ and $L_y = 10$, where OBC is that the boundary along the $x$-direction is open, the EDS is marked in green. Where $u=-1$, $\gamma_g=\gamma_l=0.2$. And the spectra of coupled damping matrix $\tilde{X}_{c,\text{qwz}}$ for $C_1$ and $C_2$ where $c=0.8$. One can see that EDS is robust against the perturbation $C_1$ (c), while it's gaped out by the perturbation $C_2$(d).}
	\label{QWZ_Cspectra}
\end{figure}

From 38-fold way of NHRM, the topological classification of $\tilde{X}_{\text{qwz}}(\mathbf{k})$ in different spatial dimensions is obtained in Table.IV.
In the 2D case, only the class $\mathcal{S}_{-}$AI with $\mathbb{Z}_2$ classification is non-trivial, the other three classes are trivial.
The TRS in class $\mathcal{S}_{-}$AI is satisfied by the coupling $C_1$ while violated by $C_2$, so it would expect that the EDS might be gaped out when the coupling $C_2$ is introduced.

In Fig.\ref{QWZ_Cspectra}, the spectra of double damping matrix $\tilde{X}_{\text{qwz}}$ in the PBC and OBC are presented. Which OBC is that the boundary condition along the $x$-direction is open. This is because that the imaginary momentum along the $y$-direction will induce the NH skin effect, that forces edge-modes of $X_{\text{qwz}}$ becomes pure imaginary, such that the EDS is absent for such boundary condition.
Furthermore, The EDS are gaped out by coupling $C_2$ in Fig.\ref{QWZ_Cspectra}(d), while they are robust to the coupling $C_1$ in Fig.\ref{QWZ_Cspectra}(c), that is to say the EDS are protected by the TRS $U_t = \tau_x\otimes\mathbb{I}$.

\section{conclusion and discussion}
In this paper, a framework to understand the topological protection of EDS in the presence of both dissipation and internal dynamics is provided.
We make use of the 38-fold way to classify the damping dynamics with EDS in dissipative TIs, of which the dissipative dynamics of these sysytems are completely captured by a single-particle NH matrix $X$.
It turns out that the symmetry classification of $X$ with EDS is ill defined, the right classification scheme is based on the double damping matrix $\tilde{X} = \text{diag}\left( X, X^* \right)$.
In our scheme, the double damping matrix $\tilde{X}$ is classified topologically by using the 38-fold way, the edge modes that close the imaginary line-gap of $\tilde{X}$ are the EDS of $X$.
Different from previous studies of EDS in purely dissipative systems \cite{topo_diss,topo_diss2}, the EDS in this work are robust against the including of the Hamiltonian terms.
As the matter of fact, in the two explanation examples of dissipative SSH model and dissipative QWZ model, the appearance of EDS are associated with non-trivial topology of internal dynamics, such that the EDS are also a dynamic signature of topological order.

In 38-fold way, the imaginary line-gap can be defined in the system of which TRS, PHS$^\dagger$ or CS is satisfied, however these symmetries are equivalent when we flatten the spectra\cite{38-fold}.
Which means the double damping matrix $\tilde{X}$ which TRS is automatically satisfied is universal for the open free fermionic system.
So it can be also applied to the dissipative TSs, the only difference is that the quansi-particle is changed into the Majorana fermions, and the EDS become the Majorana zero-damping modes, that is vital for the dissipative braiding\cite{topo_diss}.
We expect our proposal can also be used in the dissipative TSs, which is our future direction. 

\section{acknowledgment}

\end{document}